# The AMADEUS project at DAΦNE

*Catalina. Curceanu*

LNF – INFN, Via. E. Fermi 40, 00044 Frascati (Roma), Italy

On behalf on the AMADEUS Collaboration

## 1. The AMADEUS scientific case

The change of the hadron masses and hadron interactions in the nuclear medium and the structure of cold dense hadronic matter are hot topics of hadron physics today. These important, yet unsolved, problems will be the research field of AMADEUS [1] (Antikaonic Matter At DAFNE: Experiments with Unraveling Spectroscopy).

AMADEUS will search for antikaon-mediated deeply bound nuclear states, which could represent, indeed, the *ideal conditions* for investigating the way in which the spontaneous and explicit chiral symmetry breaking pattern of low-energy QCD occur in the nuclear environment.

In a few-body nuclear system the isospin I=0 $\overline{K}$N interaction plays an important role: it can favor the existence of discrete nuclear bound states of $\overline{K}$ in nuclei, while contracting the nucleus, thus producing a cold dense nuclear system, named "deeply bound kaonic nucleus" or "kaonic nuclear cluster".

Many important impacts then follow:

- such compact exotic nuclear systems might get formed with binding energies so large (~100 MeV) that their widths turn out rather narrow, since the $\Sigma\pi$ decay channel is energetically closed and, additionally, the $\Lambda\pi$ channel is forbidden by isospin selection rule;

- high-density cold nuclear matter might be formed around K⁻, which could provide information concerning a modification of the kaon mass and of the $\overline{K}$ N interaction in the nuclear medium;

- empirical information could be obtained on whether kaon condensation can occur in nuclear matter, with implications in astrophysics: neutron stars, strange stars.

- nuclear dynamics under extreme conditions (nuclear compressibility, etc) could be investigated.

The hypothesis of the possible existence of deeply bound kaonic nuclear states was already formulated in 1986 by Wycech [2] but is only a few years old in the structured form of a phenomenological model formulated by Akaishi and Yamazaki [3]. The existence or not of such deeply bound systems is presently matter of vivid discussions among theoreticians and experimentalists.

First experimental indications have been produced at KEK [4-6], LNF [7], GSI [8] and BNL [9]. Lately, the experimental situation became more complicated, since one of the two systems seen at KEK by E471, the S0(3115), was not re-seen in E570 with a larger statistics, as preliminary presented at the IX International Conference on Hypernuclear and Strange Particle Physics, Mainz 10-14 October 2006. In the same time, the interpretation of the experimental results is matter of lively discussion, as resulting from the talks given at the previously mentioned Workshop.

The new proposal, AMADEUS at DAΦNE, has the goal to perform, for the first time, a systematic and complete spectroscopic study of deeply bound kaonic nuclei, both in formation and in the decay processes.

Moreover, AMADEUS aims to perform other types of measurements as: elastic and inelastic kaon interactions on various nuclei, obtaining important information for a better understanding of the undergoing processes.

## 2. AMADEUS at DAΦNE

The ɸ-factory DAΦNE at LNF is a double-ring $e^+e^-$ collider, designed to operate at the center of mass energy of the ɸ-meson, whose decay delivers almost monochromatic K+K- pairs with a momentum of 127 MeV/c. A peak luminosity in excess of $1 \times 10^{32}$ cm$^{-2}$ s$^{-1}$ is presently obtained in the FINUDA run. Such a source of kaons is intrinsically clean, a situation unattainable with a hadron machine, which normally has an intense pion background. The planned upgrade of DAΦNE will reach a peak luminosity in excess of $10^{33}$ cm$^{-2}$ s$^{-1}$ [10], which, of course, makes it an ideal machine to study stopped K$^-$ induced reactions to search for deeply bound antikaon nuclear clusters.

This new facility will deliver an integrated luminosity of about 10 fb$^{-1}$ per year and equipped with a dedicated 4π detector KLOE [11] complemented with the AMADEUS apparatus will become the top level scientific center to study kaonic nuclei using K- induced processes at rest, as proposed in the original work of Akaishi and Yamazaki [3]. This represents a complementary approach to other experimental studies, planned at GSI with the FOPI detector and at J-PARC, E15 experiment, to be seen as part of a global strategy to attack these important open problems of low-energy QCD.

In Figure 1 the location of AMADEUS setup within KLOE detector is shown.

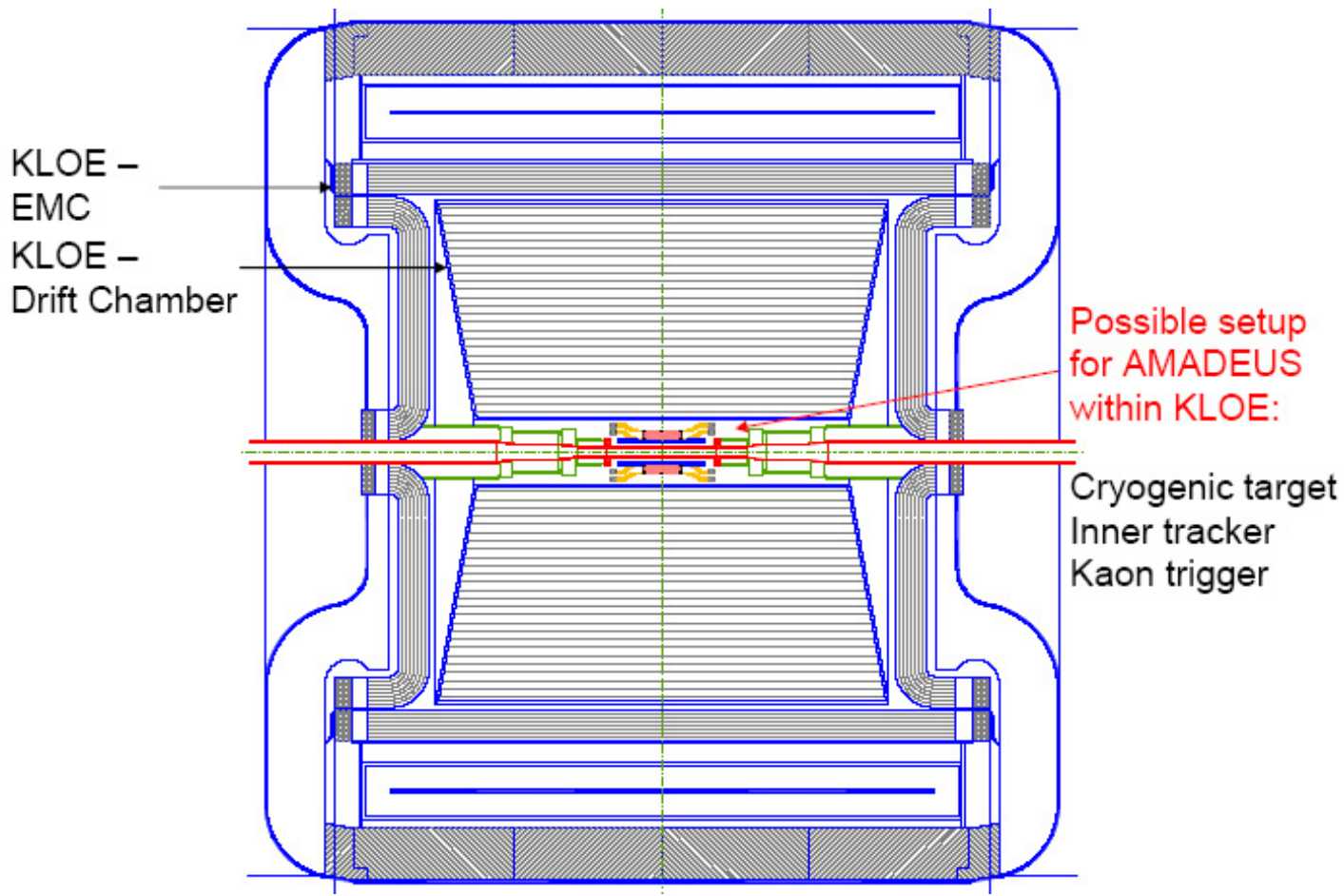

Figure 1: The AMADEUS setup within KLOE.

For the integration of the AMADEUS setup within KLOE a solution which is presently under study is to use a toroidal target placed around the beam pipe and surrounding the interaction region. The beam pipe can be a thin-walled aluminum pipe with carbon fiber reinforcement. A degrader which might be an "active" one, i.e. a scintillator (or scintillating fiber) detector is placed around the pipe just in front of the target. This detector is essential, delivering an optimal trigger condition by making use of the back-to-back topology of the kaons generated from the $\phi$-decay. The AMADEUS collaboration considers as well the implementation of an inner tracker, to get more information about the formation region (for better background suppression) of the deeply bound states. Since the

KLOE collaboration envisages the use of an inner vertex detector this might open the way to a unique inner vertex detector to be used by both groups. The possible setup is shown in Figure 2.

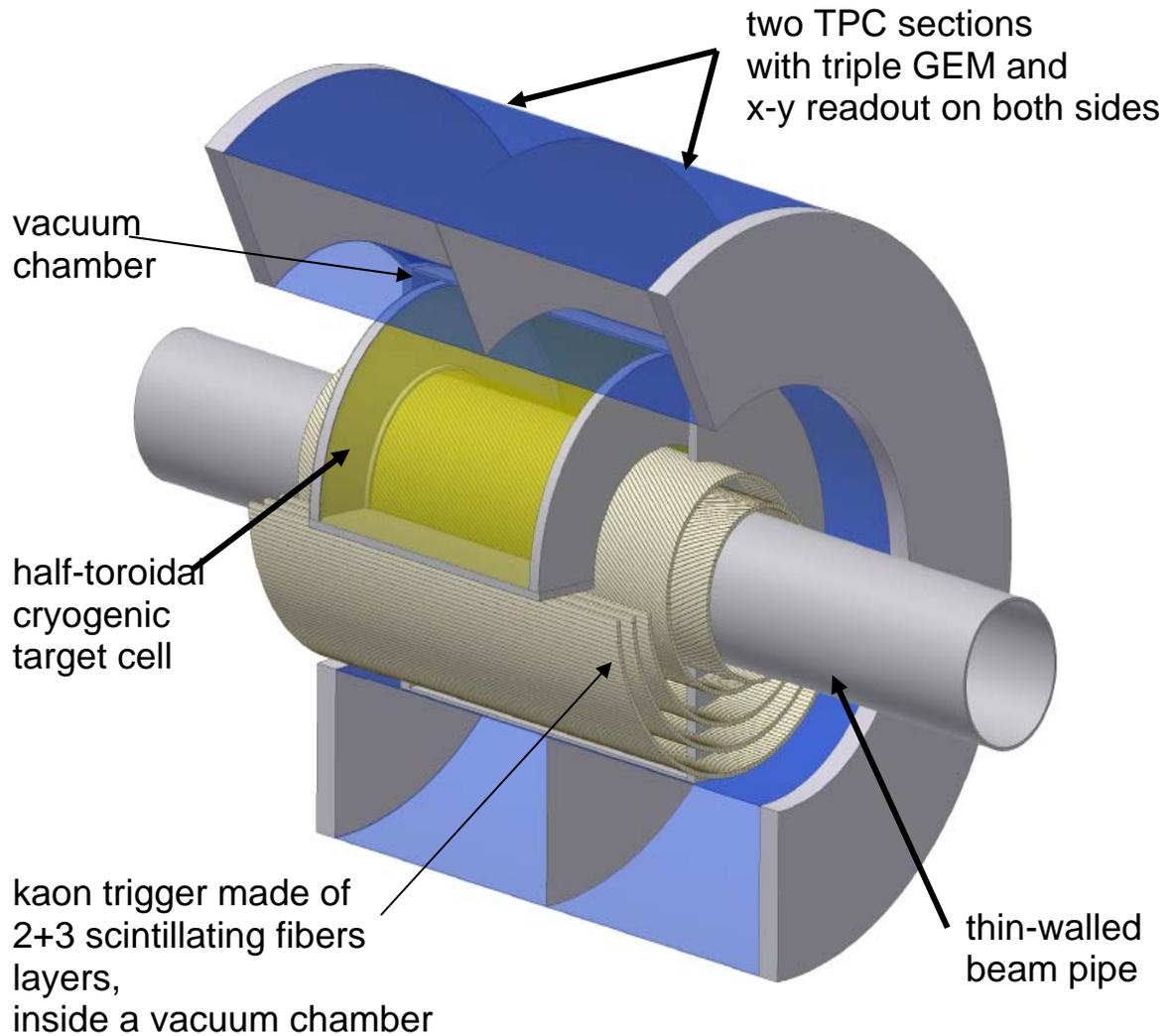

Figure 2: A possible layout of AMADEUS within KLOE

## 3. AMADEUS scientific program

The AMADEUS first phase program foresees the investigation of the most basic antikaon - mediated clusters, namely:

- kaonic dibaryon states $ppK^-$ and $pnK^-$, produced via $^3$He (stopped K, n/p) reactions;

- kaonic 3-baryon states $ppnK^-$ and $pnnK^-$, produced via $^4$He (stopped K-, n/p) reactions.

The first goal of AMADEUS is to clearly settle the question about the existence of antikaon-mediated bound nuclear clusters. To this aim a systematic and complete campaign of spectroscopic studies on light and medium nuclei looking for nuclear clusters both in formation and in decay will be performed

The program is based on the following objectives:

(i) To perform precision measurements in order to determine the quantum numbers (spin, parity, isospin) of all states, including excited ones, in addition to their binding energies and decay widths. Masses of kaonic clusters are obtained by missing mass analysis, measuring proton and neutron distributions from the formation process. In KLOE the proton spectra can be obtained with 1-2 MeV precision. For the neutron spectra the achieved precision is of the order of 2-4 MeV.

(ii) As all the states of kaonic nuclei are quasi-stationary, important information on their structure is contained in their total and partial decay widths. Until now, the experimental values on the total decay width are under discussion and no information on partial decay channels is available. Total decay widths, accessible via the formation process, can be resolved in AMADEUS at the 1-3 MeV level for proton spectra, and at a few MeV, still under study, for neutron spectra. The partial decay widths may be resolved at the level from 2 to 10 MeV depending on the decay channel.

(iii) Detailed structure information can be extracted from a Dalitz analysis of three-body decays of kaonic nuclei, as was pointed out recently by Kienle, Akaishi and Yamazaki [6].

*Acknowledgements*

We gratefully acknowledge the very good cooperation with the DAΦNE and KLOE teams.

Part of this work was supported by "Transnational access to Research Infrastructure", TARI – LNF activity within the HadronPhysics I3 project, Contract No. RII3-CT-2004-506078.